\documentclass{aa}

\usepackage{amsmath}
\usepackage{amssymb}
\usepackage{graphicx}
\usepackage{epsfig}
\usepackage{changebar}
\usepackage{natbib}
\usepackage{multirow}
\usepackage{subfigure}
\usepackage{txfonts}

\usepackage[figuresright]{rotating}

\newcommand{\kms}{km~s$^{-1}$}
\newcommand{\cmthree}{cm$^{-3}$}

\newcommand{\ks}{$K_{\rm{s}}$}

\renewcommand{\footnoterule}{ }

\begin{document}
   \title{A tale of two cores: Triggered massive star formation in the bright-rimmed cloud SFO~75}

  %\subtitle{Triggered massive star formation}

   \author{J.~S.~Urquhart
          \inst{1}
          \and
			 M.~A.~Thompson
			 \inst{2}
			 \and
			 L.~K.~Morgan
			 \inst{3}
			 \and
			 M.~R.~Pestalozzi
			 \inst{2}
			 \and
			 Glenn~J.~White
			 \inst{4,5}
			 \and
			 D.~N.~Muna
			 \inst{6}
          }

   \offprints{J. S. Urquhart: jsu@ast.leeds.ac.uk}

   \institute{School of Physics and Astronomy, University of Leeds, Leeds, LS2~9JT, UK 
         \and
             Centre for Astrophysics Research, Science and Technology Research Institute,
University of Hertfordshire, College Lane, Hatfield, AL10 9AB, UK
\and
Green Bank Telescope, P.O. Box 2, Green Bank, WV 24944, USA
\and
Dept. of Physics \& Astronomy, The Open University, Walton Hall, Milton
Keynes, MK7 6AA, UK
\and
Space Physics Division, Space Science \& Technology Division, CCLRC
Rutherford Appleton Laboratory, Chilton, Didcot, Oxfordshire, OX11 0QX,
UK
\and
Dept. of Physics \& Astronomy, University of Sheffield, Hicks Building, Hounsfield Road, Sheffield, S3 7RH, UK
             }

   \date{}

% \abstract{}{}{}{}{} 
% 5 {} token are mandatory
\abstract
   %context heading (optional)
   {Bright-rimmed clouds (BRCs) are isolated molecular clouds located on the edges of evolved HII regions. Star formation within the BRCs may have
been triggered through the propagation of photoionisation-induced shocks driven by the expansion of the HII region.}   
  % aims heading (mandatory)
   {The main focus of this paper is to investigate the current level of star formation within one of these clouds and evaluate to what extent, if any,
star formation may have been triggered.} 
  % methods heading (mandatory)
   {We present a detailed multi-wavelength study of the BRC SFO~75, including 1.3~cm and 1.2~mm continuum, and $^{13}$CO and ammonia spectral line  observations. To build up a comprehensive picture of the local environment we complement our observations with  archival data from the 2MASS, GLIMPSE and IRAS surveys.}
  % results heading (mandatory)
   {The $^{13}$CO and 1.2~mm emission reveals the presence of a dense
core located behind the bright rim of the cloud which is approximately coincident with that of the IRAS point source. From an analysis of the IRAS and 
1.2~mm fluxes we derive a dust temperature of $\sim30$~K, a luminosity of $L_{\rm{bol}}=1.6\times10^4$~L$_\odot$ and estimate the core mass to be $\sim 570$~M$_\odot$. The higher resolution ammonia observations resolve the 1.2~mm core into two distinct cores, one directly behind the cloud's rim (Core~A) and the second located slightly farther back (Core~B). These have masses of 8--15 M$_\odot$ and 3.5--7~M$_\odot$ for Core~A and Core~B respectively, which are significantly larger then their virial masses.  Comparing the morphology of Core~A with that of the photon-dominated region and ionised boundary layer leaves little doubt that it is being strongly affected by the ionisation front. 
2MASS and GLIMPSE archive data which reveal a small cluster of three deeply embedded ($A_{\rm{v}}\sim 20$~mag) high- and intermediate-mass young stellar objects towards Core~A leads us to conclude that the star formation found towards this core has been triggered. In stark contrast, Core~B appears to have a much simpler, almost spherical, morphology. No stars are found towards Core~B. We find evidence supporting the presence of shocked gas within the surface layers of the cloud which appears to extend to midway between the two ammonia cores.}
  {The scenario that emerges from our analysis is one where the two ammonia cores pre-date the arrival of the ionisation front. Since its arrival the over-pressure of the ionised gas at the surface of the cloud has driven shocks into the surface layers of the cloud. The propagation of these shocks through Core~A have triggered the formation of a small cluster of massive stars, however, the shock front has not yet propagated deeply enough into the cloud to have affected the evolution of Core~B.}

\authorrunning{J. S. Urquhart et al.}
\titlerunning{Triggered star formation within SFO 75}
\maketitle
%\tableofcontents
\section{Introduction}

Bright-rimmed clouds (BRCs) are isolated molecular clouds located on the edges of evolved HII regions. Star formation within the BRCs may thus have been
triggered through the propagation of photoionisation-induced shocks driven by the expansion of their  HII regions. The photoionisation of a BRC's surface
layers by UV photons from nearby OB stars leads to the formation of a layer of hot ionised gas. This is known as the \emph{ionised boundary layer} (IBL), which
surrounds the rim of the molecular cloud. Shocks are driven into the molecular gas if the IBL is over-pressured with respect to the molecular gas within the BRC, resulting in the compression of the cloud. This can lead to the formation of dense cores, which are then triggered to collapse by the
same (or a subsequent) shock front (\citealt{elmegreen1992}). This propagating shock front may also serve to trigger the collapse of pre-existing dense
cores, leading to the creation of a new generation of stars. 

This mode of star formation is known as \emph{radiative--driven implosion} (RDI) and may be responsible for the production of hundreds of
stars in each HII region (\citealt{ogura2002}), perhaps even contributing up to $\sim 15$ \% of the low-to-intermediate mass \emph{initial mass function}
(\citealt{sugitani2000}). In order to evaluate this mode of triggered star formation \citet{sugitani1991} and \citet{sugitani1994} identified 89 BRCs
(commonly referred to as the SFO catalogue) where star formation is likely to be taking place. These clouds were identified by correlating IRAS point sources -- having colours
consistent with embedded protostars -- with clouds displaying optically bright rims from the Sharpless HII region catalogue (\citealt{sharpless1959})
and the ESO(R) Southern Hemisphere Atlas. 

We have reported in an earlier paper the results of a set of radio continuum observations made towards all 45 SFO objects located in the southern hemisphere
(see \citealt{thompson2004} for details; hereafter Paper~I). In this paper we identified 18 BRCs that are associated with centimetre continuum emission
which is spatially and morphologically consistent with the presence of an IBL (i.e.,~radio continuum emission is elongated parallel to the optical
emission and lies in front of the cloud with respect to the direction of ionisation). Follow-up molecular line and high resolution radio continuum
observations were carried out on a sample of four clouds from Paper I in order to evaluate whether the presence of the ionised gas was having a direct
influence on the evolution of these clouds and the inferred star formation within them.  Dense molecular cores were found to be located
directly behind the ionised rims with respect to the direction of their ionising stars (\citealt{urquhart2006a}; hereafter Paper~II). Evaluating the
pressure balance between the ionised and molecular gas, we identified two clouds in approximate pressure balance and therefore likely to be in a
post-shocked state, and two clouds that are in the early stages of radiative-driven collapse with shocks still being driven into them. 

From Paper II we were able to identify two clouds at
an early stage of interaction with the expanding ionisation front and found
tentative evidence for star formation within these clouds. In this, the first of two papers, we present a
detailed multi-wavelength study of SFO~75, with SFO~76 being the subject of a subsequent paper (Urquhart et al. 2007, in prep.). These include radio and millimetre continuum, $^{13}$CO and ammonia spectral line observations as well as 2MASS, GLIMPSE and IRAS archival data. The main focus is to investigate the current level of star formation within this cloud and evaluate to what extent,
if any, the star formation may have been triggered. The structure of this paper is as follows: in Sect.~2 we summarise the main features of SFO~75
and its host HII~region; in Sect.~3 we describe the observational procedures and data reduction processes; we present the observational results and
analyses in Sect.~4; our main findings are discussed in Sect.~5 with particular emphasis on whether star formation has been triggered; and in Sect.~6 we
present a summary of our results and highlight our main findings. 

\section{SFO~75 and its environment: H$_\alpha$ and 8~$\mu$m}

SFO~75 is situated on the edge of the HII region RCW 98 that is located at a heliocentric distance of 2.8 kpc. The HII region is ionised by LSS~3423,
an O9.5 IV star (\citealt{yamaguchi1999}) which lies at a projected distance of $\sim0.8$~pc from the optically bright-rim of SFO~75. Embedded within this
cloud is the IRAS point source 15519--5430, which is the most luminous in the SFO catalogue with an far-IR luminosity (\emph{L}$_{\rm{FIR}}$) of
$\sim3.4\times10^{4}$~L$_\odot$ (\citealt{sugitani1994}).

\begin{figure*}
\begin{center}
\includegraphics[width=0.99\linewidth,trim=20 0 80 0]{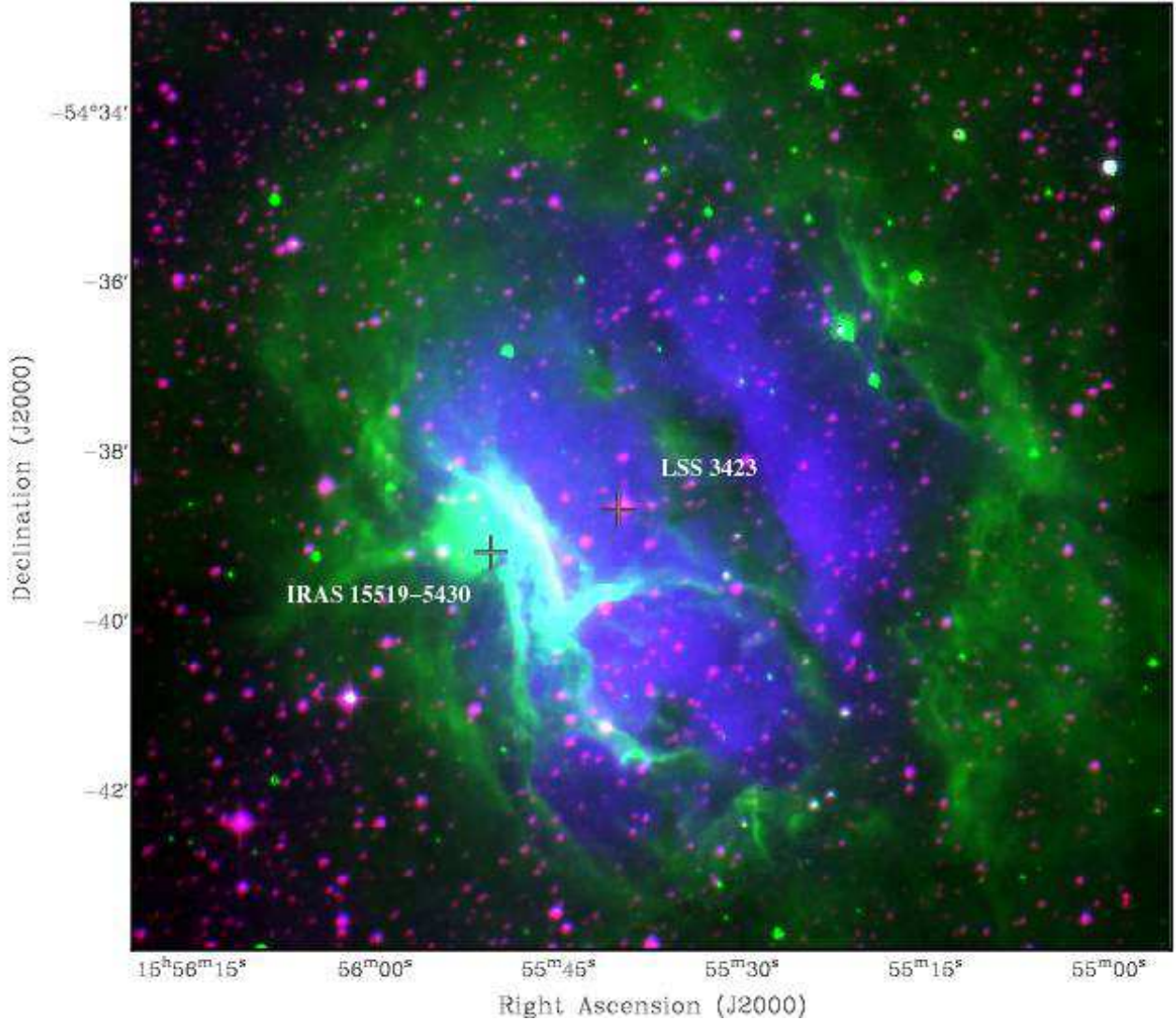}\\
\vspace{5pt}
\caption{\label{fig:sfo75_3colour_region} Three colour composite image of SFO~75 and its associated HII region, RCW 98. The position of the IRAS point source and ionising star are indicated by a black and white cross. This composite image has been produced using a combination of the DSS red band optical image, SuperCOSMOS H$_\alpha$ image and GLIMPSE 8~$\mu$m image which are coloured red, blue and green respectively.}
\end{center}
\end{figure*}

In Fig.~\ref{fig:sfo75_3colour_region} we present a three colour composite image of SFO~75 and the HII region within which it resides. This image has
been constructed from an SuperCOSMOS H$_\alpha$ image (blue; \citealt{parker2005}), optical DSS-POSS II image (red) and the Spitzer-IRAC 8~$\mu$m
(green) image obtained from the GLIMPSE survey (\citealt{benjamin2003}). The H$_\alpha$ emission arises from the recombination of ionised hydrogen
within the HII region. The 8~$\mu$m band is dominated by emission from two polycyclic aromatic hydrocarbon (PAH) features centred at 7.7~$\mu$m and 8.6~$\mu$m. PAHs 
are excited by UV photons that manage to penetrate some way into the molecular material
surrounding the HII region, and thus their presence marks the interface between the ionised gas within the HII region and the surrounding molecular gas
(\citealt{leger1984}), commonly referred to as a photo-dominated region (PDR).

The 8 $\mu$m and H$_\alpha$ emission regions can clearly be seen to be anti-correlated with each other except at the position of the BRC. The 8 $\mu$m emission
forms a continuous unbroken, approximately circular, ring (green) which completely surrounds the ionised gas (blue), indicating that the HII region is
ionisation bounded. The roughly circular appearance of the HII~region would suggest that it has expanded into a fairly homogeneous, spherically symmetric,
environment. In addition to the circular ring of emission, the 8~$\mu$m emission reveals the presence of a dust lane running from the north-east to the
south-west at an angle of $\sim45$\degr, bisecting the HII~region giving it a ``coffee bean'' like appearance. 

SFO~75 stands out dramatically against the otherwise classical looking HII region. In Fig.~1 the bright rim is seen as a thin white region that curves
slightly away from the ionising star. The thin rim seen in the optical plate is consistent with the cloud and the illuminating star being located either in the same plane of the sky, or with the cloud being located slightly in the foreground. However, the cloud cannot be too much in the foreground as we would  expect to see a region of extinction in the H$_\alpha$ emission from the dense
gas within the cloud, and this is not observed. We therefore conclude that SFO~75 protrudes into the HII region and lies approximately in the same plane as the  ionising star, and thus its projected distance from the ionising star is likely to be close to its true distance from it. 

\section{Observations and data reduction procedures}
\label{sect:observations}

\subsection{ATCA centimetre continuum and ammonia observations}
\label{sect:atca_observations}

\begin{table}[!tbp]
\begin{center}
\caption{Observational parameters for the ATCA NH$_3$ and cm observations.}
\label{tbl:atca_radio_parameters}
\begin{minipage}{\linewidth}
\begin{tabular}{lcc}
\hline
\hline
Parameter& NH$_3$ (1,1) & 1.3 cm continuum\\
\hline
Rest frequency (MHz)\dotfill & 23,695 & 23,569 \\
Total bandwidth (MHz)\dotfill& 8    & 128 \\
Primary beam (arcmin)\dotfill & 2.5 & 2.5 \\
Number of channels\dotfill   & 512  & 32 \\
Channel resolution (kHz)\dotfill & 15.6 & ---\\
Velocity resolution (km s$^{-1}$)\dotfill & 0.2 & ---\\
Date of observations\dotfill & \multicolumn{2}{c}{29 April 2004}\\
Total integration time (hr)\dotfill & \multicolumn{2}{c}{12}\\
Phase centre\dotfill & RA(J2000) & Dec.(J2000)\\
\dotfill		& 15$^h~55^m~49^s$ & -54$^\circ$~39\arcmin~13\arcsec \\
Array configurations \dotfill& \multicolumn{2}{c}{EW357}\\
Flux density calibrator\dotfill & \multicolumn{2}{c}{Mars}  \\
Bandpass calibrator\dotfill & \multicolumn{2}{c}{1253-055}  \\
Phase calibrator\dotfill & \multicolumn{2}{c}{1613--586}  \\
\hline
\end{tabular}\\
\\
\end{minipage}
\end{center}
\end{table}

Observations of the ammonia (1,1) inversion transition and the 1.3~cm radio continuum were made on the 29th April 2004 using the Australia Telescope
Compact Array (ATCA), which is located at the Paul Wild Observatory, Narrabri, New South Wales, Australia.\footnote{The ATCA is funded by the
Commonwealth of Australia for operation as a National Facility managed by CSIRO.} The ATCA consists of $6\times22$~m antennas, 5 of which lie on a 3~km
east-west railway track with the sixth antenna located 3~km farther west. This allows the antennas to be positioned  in several configurations with
maximum and minimum baselines of 6~km and 30~m respectively. 

The observations were made at two different frequency bands centred at 23,569 and 23,695 MHz (NH$_3$ (1,1))  using bandwidths of 128 and 8 MHz
respectively. The 8 MHz bandwidth was divided into 512 channels, giving a frequency resolution of 15.6 kHz and a velocity resolution of
0.2~km~s$^{-1}$ per channel. The observations were made over a twelve hour period providing good hour angle coverage. To correct for fluctuations in the phase and
amplitude caused by atmospheric and instrumental effects the total integration time was split into blocks of $\sim$ 15 minutes of on-source integration
time sandwiched between two minute observations of the phase calibrator 1613$-$586. For absolute calibration of the flux density Mars was observed once
during the observations for approximately ten minutes. To calibrate the bandpass the bright point source 1253$-$055 was also observed once during the observations. The main observational parameters are summarised in Table~\ref{tbl:atca_radio_parameters}.

The calibration and reduction of the data were performed using the MIRIAD reduction package \citep{sault1995} following standard ATCA procedures. The
task UVLIN was used to subtract the continuum emission in the 23,695 MHz bandpass from the ammonia (1,1) spectra in the \emph{uv} plane. These data
were then CLEANed using a robust weighting of 0.5 to obtain the same sensitivity as natural weighting, but with a much improved beam-shape and lower
sidelobe contamination.  The CLEANed data were restored to produce an image data cube of the ammonia (1,1) emission. An image of the 1.3 cm radio
continuum emission was produced from the second frequency bandpass.

\subsection{Mopra $^{13}$CO observations}
\label{sect:co_observations}

In Paper II we presented CO observations made in June 2003 towards SFO~75 with the Mopra telescope. Mopra is a 22 metre telescope located near
Coonabarabran, New South Wales, Australia.\footnote{Mopra is operated by the Australia Telescope National Facility, CSIRO and the University of New
South Wales.} The telescope is situated at an elevation of 866 metres above sea level, and  at a latitude of 31 degrees south. 

Emission maps were constructed from a number of single pointing observations made over a grid of positions in the $^{12}$CO, $^{13}$CO and
C$^{18}$O ($J$=1--0) rotational lines. These observations revealed the presence of an embedded core within the rim, and allowed us
to estimate its physical parameters (e.g., excitation temperature, density, mass), however, these maps proved too small ($2\arcmin\times2\arcmin$) to study the
kinematics of the molecular gas. We therefore re-observed this cloud in July 2005 using the recently commissioned \emph{on-the-fly} mapping mode
to map the distribution of $^{13}$CO in a $5\arcmin\times5\arcmin$ region centred on the embedded core identified from our previous observations. 

\begin{table}[!tbp]
\begin{center}
\caption{Observational parameters for the Mopra $^{13}$CO observations.}
\label{tbl:CO_radio_parameters}
\begin{minipage}{\linewidth}
\begin{tabular}{lcc}
\hline
\hline
Parameter& \\
\hline
Rest frequency (GHz)\dotfill & \multicolumn{2}{c}{110.201} \\
Total bandwidth (MHz)\dotfill& \multicolumn{2}{c}{64} \\
Number of channels\dotfill   & \multicolumn{2}{c}{1024} \\
Velocity resolution (km s$^{-1}$)\dotfill & \multicolumn{2}{c}{0.17}\\
Beam size (\arcsec)\dotfill & \multicolumn{2}{c}{34} \\
Date of observations\dotfill & \multicolumn{2}{c}{7th July 2005}\\
Total integration time (hr)\dotfill & \multicolumn{2}{c}{$\sim$1.5}\\
Obs. positions\dotfill & RA(J2000) & Dec.(J2000)\\
Map centre\dotfill		& 15$^h~55^m~49.3^s$ & -54$^\circ$~39\arcmin~13\arcsec \\
Offset position\dotfill		& 16$^h~02^m~17.6^s$ & -55$^\circ$~19\arcmin~12\arcsec \\
\hline
\end{tabular}\\
\\
\end{minipage}
\end{center}
\end{table}

The receiver is a cryogenically cooled \mbox{($\sim$ 4 K)}, low-noise, Superconductor-Insulator-Superconductor (SIS) junction mixer with a frequency
range between 85--116 GHz, corresponding to a half-power beam-width of \mbox{36\arcsec--33\arcsec} (Mopra Technical Summary\footnote{Available at
http://www.narrabri.atnf.csiro.au/mopra/.}). The receiver can be tuned to either single or double side-band mode. The incoming signal is separated into
two channels using a polarisation splitter, each of which can be tuned separately allowing two channels to be observed simultaneously. The receiver
backend is a digital autocorrelator capable of providing two simultaneous outputs with an instantaneous bandwidth between 4--256 MHz.

The observational setup for our $^{13}$CO map is summarised in Table~\ref{tbl:CO_radio_parameters}. For observations of SFO 75 both polarisations were
tuned to the $^{13}$CO frequency, with the the second channel being periodically tuned to \mbox{86.2 GHz} (SiO maser frequency) to check the telescope pointing at
hourly intervals. The pointing offsets were found to be $\le10^{\prime\prime}$ rms. A total of 35 rows were scanned to produce the map with an
off-source position being observed at the completion of each row to allow subtraction of sky emission. Each row was scanned at a speed of
$\sim3.2$~arcsec~s$^{-1}$ with the data being averaged over a 2 second cycle time. Individual rows were separated by 9\arcsec. To correct the
measured antenna temperatures ($T^*_{\rm{A}}$) for atmospheric absorption, ohmic losses and rearward spillover, a measurement was made of an ambient
load (assumed to be at 290~K) following the method of \citet{kutner1981} approximately every 15 minutes. System temperatures were found to be stable
during the observations; $T_{\rm{sys}} \sim 500$~K with variations of no more than 10\% during the map of SFO 75. Absolute calibration was performed by
comparing measured line temperatures of Orion KL and M17SW to standard values. We estimate the combined calibration uncertainties to be no more than
20\%.

The data were reduced using LIVEDATA and GRIDZILLA packages available from the ATNF. LIVEDATA performs a bandpass calibration for each row using the off-source data followed by fitting a user-specified polynomial to the spectral baseline. GRIDZILLA grids the data to a user-specified weighting and beam parameter input. The data were weighted using the periodic $T_{\rm{sys}}$ measurements and a cell size of 12\arcsec\ was used to grid the data.

\subsection{SEST millimetre continuum observations}

SFO~75 and its host HII region were observed in August 2002 using the 37-channel bolometer array SIMBA (SEST Imaging Bolometer Array) at the Swedish
ESO Submillimetre Telescope (SEST). SIMBA operates at a central frequency of 250~GHz which corresponds to a wavelength of 1.2~mm, with a main beam
efficiency of 0.5 and bandwidth of 50~GHz. The beam has a half power beam width of $\sim24$\arcsec\ for a single element with a sky separation of
44\arcsec\ between elements.\footnote{For more details see http://www.ls.eso.org/lasilla/Telescopes/SEST /html/telescope-instruments/simba/index.html.}

An area of approximately 800~arcsec$^2$ surrounding the IRAS point source associated with SFO~75 was mapped using the fast mapping mode
of SIMBA (for details see \citealt{Weferling2002,Reichertz2001}). Two maps of 30 minutes' duration and with a scan speed of 80 arcsec s$^{-1}$ were coadded to make the final map. The data were reduced
following the standard procedures of the MOPSI reduction package, which was written by Robert Zylka (Grenoble Observatory)\footnote{See also
\texttt{http://www.astro.ruhr-uni-bochum.de/nielbock/\break simba/mopsi.pdf}}. Atmospheric extinction was measured by skydips before and after each map, with
zenith opacities at 225 GHz of typically 0.3. Pointing was checked on an hourly basis and was found to be within 5$^{\prime\prime}$. Absolute flux
calibration was carried out by measurements of the 1.2 mm flux of Uranus. The final rms noise level in the map was found to be $\sim20$~mJy~beam$^{-1}$
and the calibration uncertainty is estimated to be $\sim10$\%.

\section{Results and analysis}
\label{sect:results}

\subsection{Image analysis}
\label{sect:image_results}

\begin{sidewaysfigure*}
\begin{center}

\includegraphics[width=0.49\linewidth, trim=0 50 0 20]{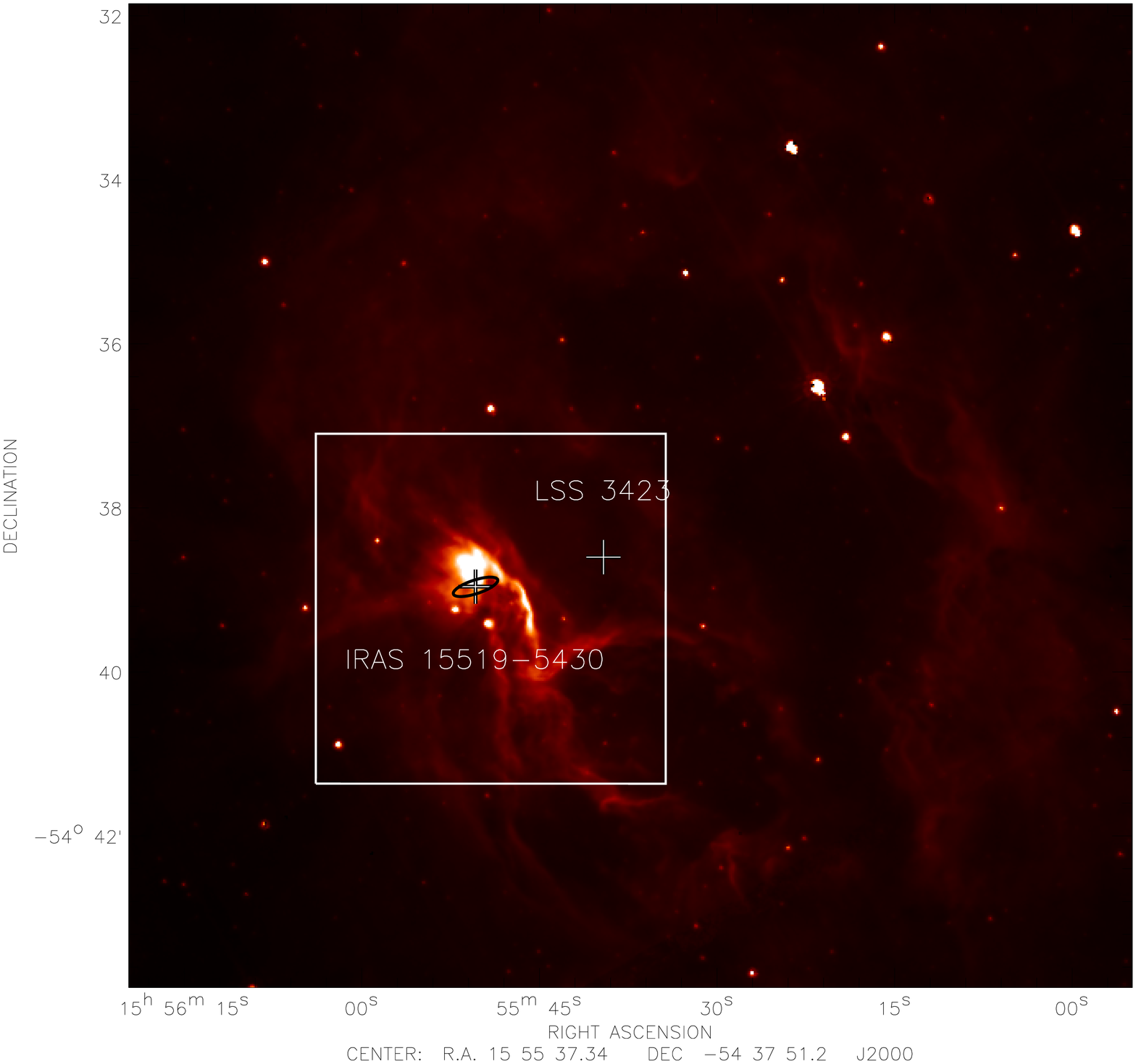}
\includegraphics[width=0.49\linewidth, trim= 0 0 0 0]{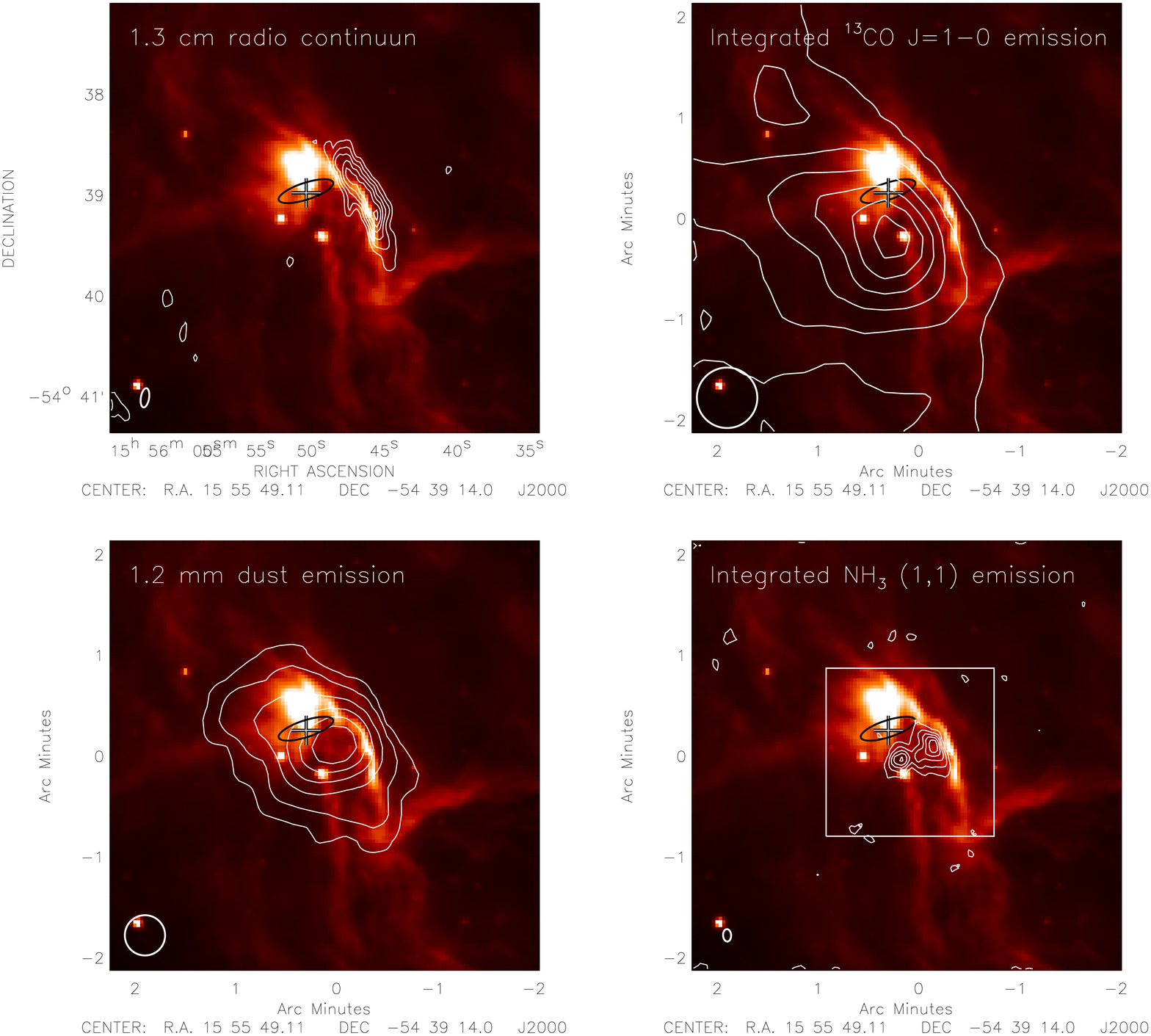} 

\caption{\label{fig:obs_images} Left panel: Large scale 8 $\mu$m GLIMPSE image of SFO 75 and its associated HII region, RCW 98. The position of the IRAS point source and ionising star are indicated by a white cross. At 8 $\mu$m the emission is dominated by PAH emission that are found in photo-dissociation region and marks the interface between the ionised gas within the HII region and the surrounding molecular gas. Right panels: Observational results laid over the region of the 8 $\mu$m GLIMPSE image outlined in white in the left panel. Clockwise from the top left panel: 1.3~cm radio continuum, integrated $^{13}$CO emission, integrated ammonia (1,1) emission (main line) and the 1.2~mm continuum emission. The position of the IRAS point source is indicated by a white cross and its positional uncertainty by a black ellipse.  The respective beam shape of each telescope is shown to size in the lower left corner of each plot. Additionally, in the image showing the $^{13}$CO emission we have plotted a dashed line to indicate the cut through the cloud used to produce the position-velocity diagram presented in Fig.~\ref{fig:velocity_position_diagram}. The integrated $^{13}$CO and ammonia emission has been integrated between $-$32 and $-$40 \kms, and $-$34.5 and $-$40 \kms\ velocity ranges respectively. The contour levels used in the image are as follows: 1.3~cm radio contours begin at 3$\sigma$ and increase in steps of 2$\sigma$ where $\sigma = 2.5$~mJy; $^{13}$CO contours begin at 4$\sigma$ and increase in steps of 3$\sigma$ where $\sigma = 9$~K \kms; 1.2~mm contour levels are 0.1, 0.2, 0.4, 0.7, 1.1 and 1.6~Jy; ammonia contours begin at 4$\sigma$ and increase in steps of 3$\sigma$ where $\sigma = 80$~mJy \kms.}    

\end{center}
\end{sidewaysfigure*}

We  present the images obtained from our observations in Fig.~\ref{fig:obs_images}. The centimetre continuum emission can clearly be seen to form a thin layer that is elongated parallel to the optical rim and located on the outside edge of the PDR, thus confirming the presence of an IBL between the
HII~region and the molecular cloud. The $^{13}$CO emission traces the molecular gas which is concentrated behind the bright rim falling off steeply in
the direction of the HII~region. The lowest $^{13}$CO contour is almost parallel with the major axis of the IBL and the bulk of the molecular material
is  elongated parallel to the bright rim and perpendicular to the direction of ionisation, consistent with the hypothesis that the cloud is
being compressed by the ionisation front and its associated shocks. 

The 1.2~mm continuum emission appears to be elongated parallel to the ionisation front with a compact core of bright emission located almost directly behind the optical rim. This emission arises from a combination of optically thin thermal dust emission and thermal free-free emission from ionised gas. It is therefore possible that a significant proportion of the observed emission does not arise from the presence of an embedded core but from ionised gas associated with the bright rim. However, we can estimate the contribution from ionised gas at 1.2~mm by extrapolating from the integrated 1.3~cm emission (30.3~mJy; see Paper~II for details) assuming a thermal spectral index, i.e., $\nu^{-0.1}$, which we find to be $\sim 24$~mJy. This is less than one per cent of the total integrated emission of 7.1~Jy. The contribution to the total flux from the thermal free-free emission is negligible and therefore the measured flux predominately arises from thermal dust emission. The 1.2~mm emission has a similar distribution to that of the $^{13}$CO emission, although not as extended and peaks closer to the bright rim than the $^{13}$CO emission ($\sim26$\arcsec), possibly indicating that the SIMBA
observations are sampling warmer material located closer to the rim. The
$^{13}$CO and dust emission peaks appear to be centrally condensed and are  located slightly south of the IRAS position.

The higher angular resolution of the ammonia data resolves the 1.2~mm core into two distinct components, one that is located
directly behind the bright rim (hereafter Core~A), and a second which is located farther back from the rim (hereafter Core~B) directly behind
Core~A with respect to the direction of ionisation. The face of Core~A nearest to the bright rim has a flattened morphology which almost exactly matches that of the inside edge of the PDR and the radio emission. This morphological correlation clearly indicates that
these three regions are interacting with each other and that the morphology of Core A is being directly influenced by the UV illumination of LSS 3423.
Core~B appears to have a more spherical morphology, possibly indicating that it has not so far been significantly influenced by the passage of the
photoionisation-induced shocks.

In Table 3 we present a summary of the core parameters determined from the various emission maps.

\begin{table*}
\begin{center}
\caption{Image parameters and peak positions of detected cores.}
\label{tbl:peak_positions}
\begin{minipage}{\linewidth}
\begin{tabular}{lcccccc}
\hline
\hline
Core tracer &  \multicolumn{2}{c}{Beam parameters} & \multicolumn{2}{c}{Position} &IRAS offset & Deconvolved core \\
&$maj\times min$ (\arcsec)& PA (\degr)& RA(J2000) & Dec.(J2000) &(\arcsec) & FWHM (\arcsec) \\

\hline
%IRAS PS &&&& 15$^h~55^m~50.4^s$ & -54$^\circ$~38\arcmin~58\arcsec \\
$^{13}$CO &33&$\cdots$&15$^h~55^m~50.9^s$ & $-$54$^\circ$~39\arcmin~26\arcsec &28&41\\
1.2-mm    &24&$\cdots$&15$^h~55^m~48.7^s$ & $-$54$^\circ$~39\arcmin~8.4\arcsec&18&36 \\
NH$_3$ (Core A)&$11.5\times4.9$&-9.2& 15$^h~55^m~47.5^s$ & $-$54$^\circ$~39\arcmin~10.2\arcsec&27&10.4\\
NH$_3$ (Core B)&$11.5\times4.9$&-9.2& 15$^h~55^m~49.5^s$ & $-$54$^\circ$~39\arcmin~16.6\arcsec&20&6.6 \\
\hline

%Core tracer & \multicolumn{2}{c}{Position}  & Source \\
%& RA(J2000) & Dec.(J2000)  & FWHM (\arcsec) \\

%\hline
%IRAS PS & 15$^h~55^m~50.4^s$ & -54$^\circ$~38\arcmin~58\arcsec \\
%$^{13}$CO& 15$^h~55^m~50.9^s$ & -54$^\circ$~39\arcmin~26\arcsec &56\\
%1.2-mm   & 15$^h~55^m~48.7^s$ & -54$^\circ$~39\arcmin~8.4\arcsec&36 \\
%NH$_3$ (Core A)& 15$^h~55^m~47.5^s$ & -54$^\circ$~39\arcmin~10.2\arcsec&11.4\\
%NH$_3$ (Core B)& 15$^h~55^m~49.5^s$ & -54$^\circ$~39\arcmin~16.6\arcsec&9.0 \\
%\hline
\end{tabular}\\
\\
\end{minipage}
\end{center}
\end{table*}

\subsection{Analysis of global properties of the ionised and molecular gas}
\label{sect:co_radio}

In Paper II we presented a detailed analysis of the internal and external gas found towards SFO~75 using our previous CO and 1.3~cm radio continuum observations, we therefore only provide a summary of the relevant findings here. From our CO analysis we found that SFO~75 has a velocity of $-$37.5 km~s$^{-1}$, a core kinetic temperature of $\sim24$~K, a particle density of $n = 3.4\times10^4$~cm$^{-3}$ and a core mass of $M_{\rm{{^{13}}CO}}=177$~M$_\odot$. Analysis of the ionised gas determined the peak ionising photon flux to be $260\times10^8$~cm$^{-2}$~s$^{-1}$ and electron density to be $n_{\rm{e}}$ $\sim$~839~cm$^{-3}$. We estimated the internal and external pressures and found that the pressure of the ionised gas is approximately three times higher than the internal pressure of the embedded CO core. 

\begin{figure}
\begin{center}

\includegraphics[width=\linewidth]{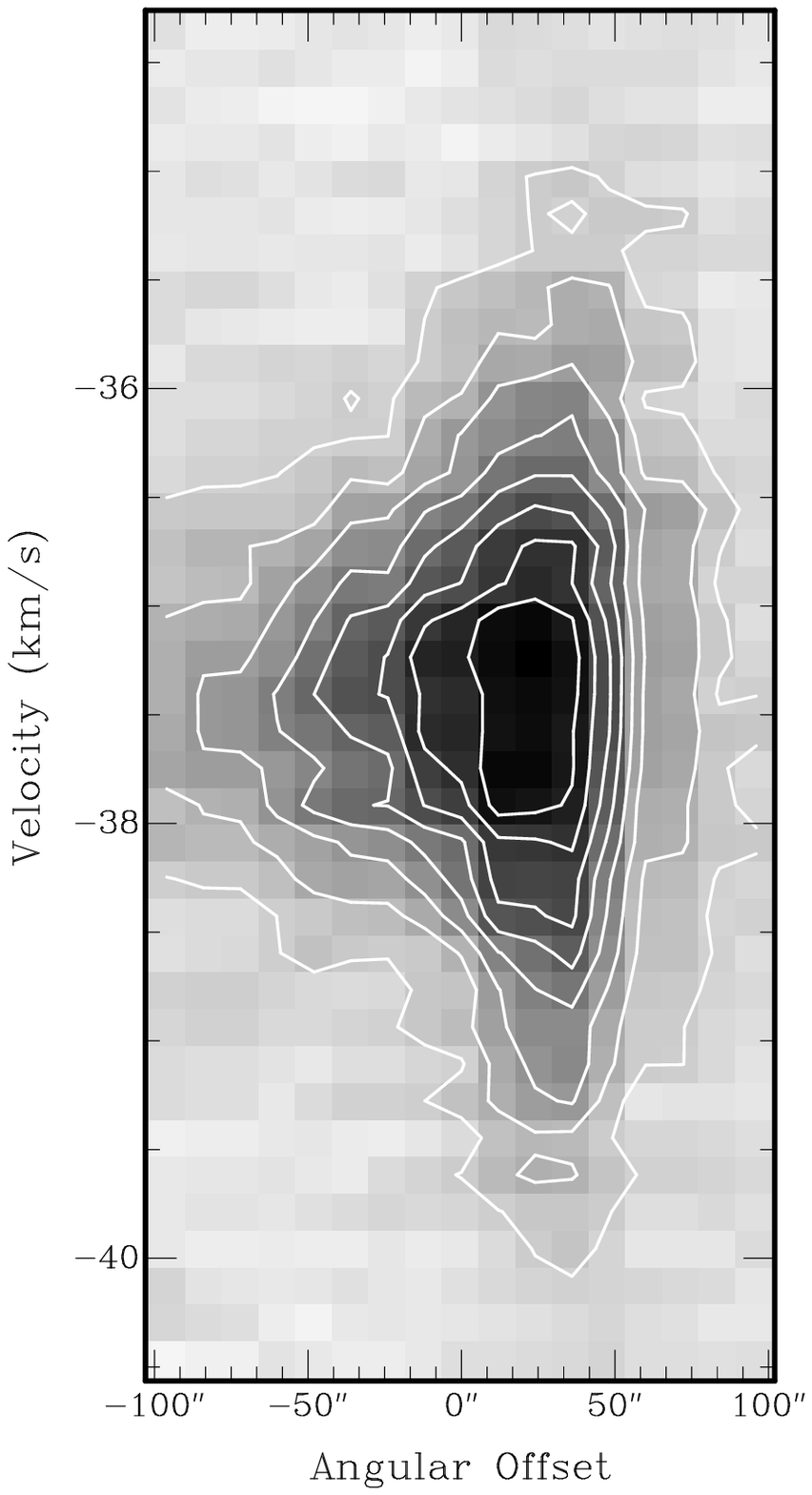}
\caption{\label{fig:velocity_position_diagram}Position-velocity diagram taken through the ionising star-CO core vector as indicated in Fig.~\ref{fig:obs_images}. The direction of ionisation is from right to left. This diagram clearly reveals the presence of blue and red shifted emission at the front of the molecular cloud, indicative of shocks and the steep emission gradient consistent with compression of the cloud by the ionisation front. The contours start at 20\% and increase in steps of 10\% of the peak value (11.1 K \kms).} 

\end{center}
\end{figure}

A drawback with our previous CO observations was the relatively small size of the maps ($2\arcmin\times2\arcmin$) and rather coarse sampling of the data which limited our analysis, particularly with respect to investigating the kinematics of the cloud. For this reason we re-observed this cloud using the $^{13}$CO~($J$=1--0) line, mapping a larger region ($5\arcmin\times5\arcmin$). The integrated emission is shown in the \emph{right panel} of Fig.~\ref{fig:obs_images}. In Fig.~\ref{fig:velocity_position_diagram} we present a position-velocity diagram taken through the cloud along the star-CO core vector (as indicated in Fig.~\ref{fig:obs_images}) starting at the position of the ionising star (north-west corner of the image) and cutting through the core, running right to left through the diagram. 

This diagram reveals the molecular material to have a steep emission gradient on the surface exposed to the ionising star, and exhibits large velocity wings ($\sim 6$~\kms) which are perpendicular to the direction of ionisation. These are indicative of shocked gas, with the blue shifted component being associated with compressed gas moving towards us and the red shifted component being associated with compressed gas moving away from us (\citealt{lefloch1994}). The similarity between the position-velocity diagram predicted by the \citet{lefloch1994} RDI model and our observations is striking (c.f. with their Fig.~14a). The presence of an IBL at the surface of the cloud, detection of a dense embedded core and the large pressure difference between the two regions led us to conclude in Paper~II that this cloud was likely to be in the early stages of radiatively driven collapse. This conclusion is strengthened with the detection of the shocked gas and the similarities between our observational results and the predictions of the RDI model. 

The shocked gas layer can be clearly distinguished from the bulk motion of gas behind by its significantly broader velocity distribution. We are therefore able to estimate how far this shocked layer extends into the cloud from the FWHM of the shocked region. We thus estimate the shocked region to be $\sim30$\arcsec, which corresponds to a physical distance of $\sim0.4$~pc, extending from the edge of the cloud's rim to just in front of the CO core. Interestingly, this would place it approximately midway between the two ammonia cores as they are seen in projection.

\subsection{1.2~mm emission analysis}
\subsubsection{Luminosity, dust temperature and mass of the mm~core}
\label{sect:mass_luminosity}

\begin{figure}
\begin{center}

\includegraphics[width=0.98\linewidth]{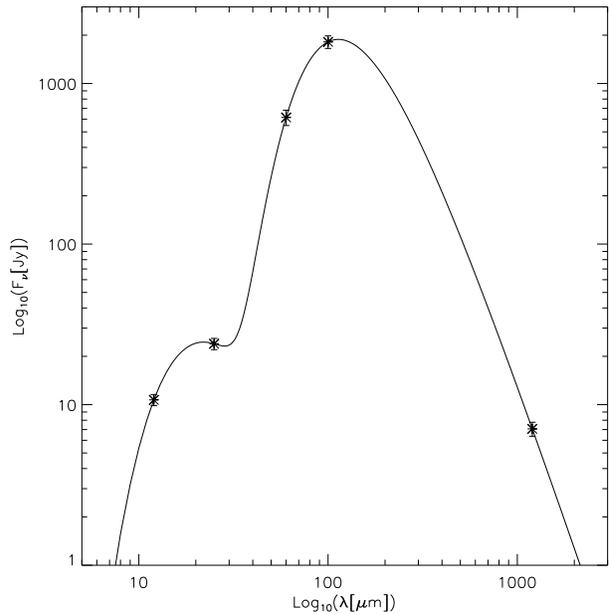}
\caption{\label{fig:sed} Spectral energy distribution of SFO~75. Results of a two component greybody fit to the IRAS and 1.2~mm dust continuum
emission.}
\end{center}

\end{figure}

In Fig.~2 we presented a contoured image of the 1.2~mm dust emission detected towards SFO~75, which has a peak flux density of 2.0~Jy beam$^{-1}$ and integrated flux of 7.1~Jy. In this subsection we use these flux measurements to derive the luminosity, temperature and mass of the embedded mm~core.

It has been shown (e.g., \citealt{sridharan2002,beuther2002}) that two components are present in the spectral energy distribution (SED) of luminous YSOs: a compact hot component which dominates the IRAS 12 $\mu$m and 25 $\mu$m fluxes, and a more extended component arising from colder gas which dominates the IRAS 60 $\mu$m and 100 $\mu$m fluxes. We therefore estimate the luminosity and temperature of these components by fitting two greybody functions to the core's SED. The greybody functions have the form (e.g., \citealt{dent1998}),

\begin{equation}
F_\nu = \Omega B_\nu(T_{\rm{dust}})(1-\rm{e}^{-\tau_\nu}),
\end{equation}

\noindent where $F_\nu$ is the flux measured at a frequency $\nu$, $\Omega$ is the solid angle subtended by the cloud, $B_\nu(T_{\rm{dust}})$ is the
Planck function evaluated at the dust temperature  ($T_{\rm{dust}}$) and frequency  ($\nu$), and $\tau_\nu$ is the optical depth at frequency $\nu$.

The SED of the 1.2 mm core was derived using fluxes from the four IRAS bands and the integrated 1.2~mm dust emission from the SIMBA map shown in
Fig.~\ref{fig:obs_images}. The optical depth $\tau$ is parameterized in terms of a common dust emissivity index $\beta$ so that $\tau$ can be evaluated at
arbitrary frequencies from a known reference frequency $\nu_{\rm{ref}}$ and optical depth $\tau_{\rm{ref}}$, i.e., $\tau_\nu=\tau_{\rm{ref}}(\nu/\nu_{\rm{ref}})^\beta$. $\beta$ has typical values of 1--2 for molecular clouds, however, to limit the number of free
parameters, and to maintain consistency with previous work of this kind (e.g., \citealt{faundez2004,sridharan2002}), we have adopted a value of 1 for the hot component, but leave the emissivity of the cold component as a free parameter. We also assume that the size of the region giving rise to the cold component emission is the same as the 1.2~mm core. We leave the temperatures of the hot and cold components, and the size of the hot component as free parameters. The two component greybody fit to the SED is presented in Fig.~\ref{fig:sed}; the temperatures for the hot and cold components are 175~K and 30~K respectively. We estimate the size of the hot central region to be $\sim 0.2$~\arcsec\ and the dust emissivity of the cold component to be $\sim1.5$. It is the cold dust temperature that is of most interest to us as it best describes the temperature of the large scale emission traced by our 1.2~mm observations. 

Integrating over the observed SED yields a bolometric luminosity of $\sim  1.6\times10^4$~L$_\odot$, which corresponds to the presence of a single ZAMS star with a spectral class between B0 and B0.5 (\citealt{panagia1973}). Alternatively, if the observed emission is due to the presence of an embedded cluster, the luminosity of the cluster's most massive member would typically be one or two subclasses lower than assuming a single star, i.e., B2 or B3 star (\citealt{wood1989}). Since the majority of the luminosity is provided by the IRAS 60 $\mu$m and 100 $\mu$m flux densities it is necessary to consider where the emission originates. Taking into account the size of the IRAS beam ($\sim2$\arcmin\ at 100 $\mu$m), which is much larger than the mm~core, it is possible that some of the measured flux arises from the ionisation front as well as the embedded core, and therefore the derived luminosity should be considered as an upper limit. In order to better constrain the SED we attempted to obtain flux density measurements from the MIPSGAL Spitzer Legacy project\footnote{For details see \texttt{http://ssc.spitzer.caltech.edu/legacy/\break abs/carey.html.}} which has recently finished surveying the GLIMPSE region at 24~$\mu$m and 70~$\mu$m. However, this particular source lies on the edge of one of the observed fields, making it impossible to extract useful flux density measurements. Given the location of the IRAS point source and its close proximity to the mm~core ($\sim18$\arcsec) we would suggest that the majority of the flux is associated with the core, and  that the core's true luminosity is close to the upper limit.

As stated previously, the majority of the millimetre emission is due to optically thin thermal dust emission and by making a few assumptions it is possible to relate the integrated millimetre flux density to the total  mass  (gas+dust) of the core via (\citealt{hildebrand1983}),

\begin{equation}
M_{\rm{(gas+dust)}} = g \frac{F_{1.2\rm{mm}}D^2}{\kappa_{1.2\rm{mm}} B_{1.2\rm{mm}}(T_{\rm{dust}})}~ \rm{M}_\odot,
\end{equation}

\noindent where $F_{1.2\rm{mm}}$ and $\kappa_{1.2\rm{mm}}$ are the integrated flux density and dust opacity per unit mass measured at 1.2~mm respectively, $B_{1.2\rm{mm}}(T_{\rm{dust}})$ is the Planck function for a dust temperature $T_{\rm{dust}}$, $D$ is the distance in kpc and $g$ is the gas-to-dust ratio, which we assume to be 100. Using a value for $\kappa_{1.2\rm{mm}}=1 ~\rm{cm}^2~g^{-1}$ (\citealt{ossenkopf1994}) and a value of 30 K for the dust temperature (as determined for the cold dust component of the greybody fit) we estimate the mass of the mm~core to be $\sim570$~M$_\odot$. Given the uncertainties involved in this calculation, and that of the CO derived mass (i.e., $M_{\rm{{^{13}CO}}}=177~$M$_\odot$), we find them to be in reasonable agreement. 

The luminosity and temperatures of the hot and cold components, which are similar to those found towards other massive embedded objects (e.g., \citealt{sridharan2002}), and the derived mass leads us to conclude  that massive star formation is currently taking place within SFO~75. This is consistent with the statistical trend for BRCs to preferentially form more massive stars or multiple stellar systems than found towards Bok globules and isolated dense cores (e.g., \citealt{sugitani1994, yamaguchi1999}).  

\subsubsection{Molecular ring surrounding the HII~region}

The only 1.2 mm emission detected in our wide-field SIMBA observations was that associated with SFO 75. This implies that the molecular shell of
material surrounding the HII region (inferred from the presence of the PDR traced by the 8 $\mu$m PAH emission) must be of relatively low column
density. We are therefore unable to determine the gas density within this ring, however, we can use the image rms to place an upper limit on the mass/beam. Assuming that the total mass is contained within a cylinder with a physical radius equal to that of the telescope beam on the sky and a length assumed to be equal to the width of the PDR seen in projection (i.e., $\sim$ 1~pc) we can estimate an upper limit for the density of the surrounding gas to be 10$^3$~\cmthree. 

This upper limit for the H$_2$ particle density in the ring implies that the expanding HII region has not yet swept up much of the surrounding material, either because the HII region is relatively
young or that the surrounding molecular gas is of low density. The low density of the molecular shell surrounding RCW 98 suggests the ``collect and collapse'' process (e.g.,~\citealt{deharveng2005,zavagno2006}) has not occurred around RCW 98 and that SFO 75 is likely to be a pre-existing molecular cloud recently exposed to the UV illumination of LSS 3423.

\subsection{Analysis of the ammonia cores}
\label{sect:ammonia_cores}

\begin{figure*}
\begin{center}
\includegraphics[width=\linewidth]{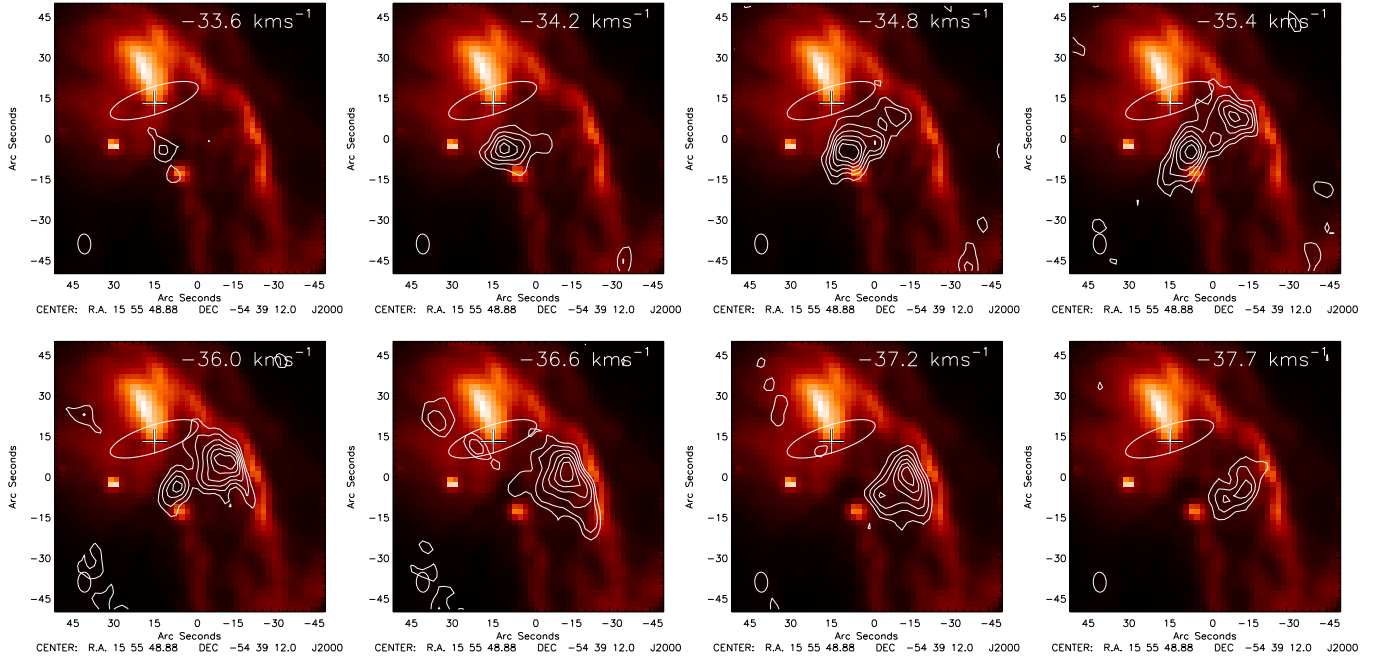} %SPHERICAL  /export/linast23_c/jsu/C1088
\caption{\label{fig:ammonia_channel_maps} Velocity channel maps of the ammonia (1,1) main line emission contoured over an enlarged region of 8~$\mu$m image presented in Fig.~\ref{fig:obs_images} (highlighted white square in lower right image). The central velocities are indicated in the top right corners of each map and are 0.6~\kms\ thick.  The position of the IRAS point source and its associated error are indicated by a cross and ellipse respectively. The first contour starts at 4$\sigma$ and increase in steps of 4$\sigma$. The size of the synthesised beam is shown to scale in the lower left corner.}
\end{center}

\end{figure*}

In Fig.~\ref{fig:ammonia_channel_maps} we present eight velocity channel maps covering the velocity range between
$-$33 to $-$38~km~s$^{-1}$. These channel maps highlight important differences between the two ammonia cores. First, the two cores are distinct both spatially and in velocity, with Core~A and Core~B
centred at approximately $-$37~\kms\ and $-$35~km~s$^{-1}$ respectively (see also Fig.~6). Secondly, they illustrate the excellent correlation of the morphology of the flattened forward face of Core~A with that of the PDR traced by the 8~$\mu$m emission -- the morphologies of these two regions leaves little doubt that the ionisation front is having a dramatic influence on the evolution of this core.
Additionally, the ends of the flattened face appear to curve into the cloud,
reminiscent of the small ``ears'' seen to develop in the \citet{lefloch1994} model (see their Fig.~4b) where the compressed layer meets the rear edge
of the initial condensation. Conversely, Core~B appears to have a more spherical morphology indicating that the shock front has not yet reached it. This is confirmed by the position of the shock front which is found between the two cores A and B (see Sect.~4.2).

\begin{figure}
\begin{center}

\includegraphics[width=\linewidth]{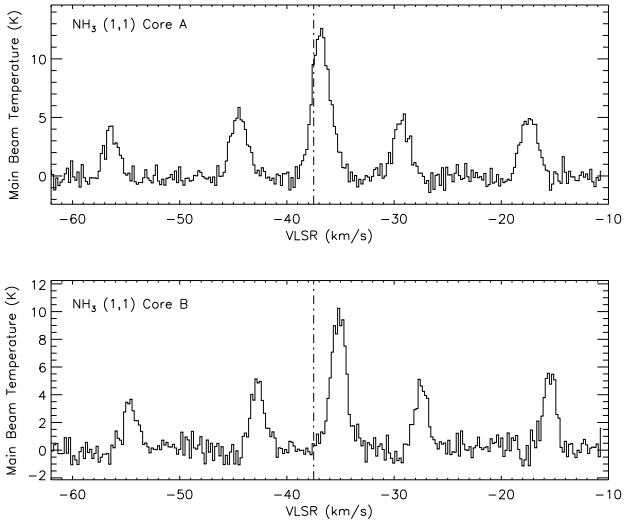}
\caption{\label{fig:ammonia_profiles} Spatially integrated spectral line profiles for Core~A and Core~B. The vertical dashed-dotted line indicated the velocity of the $^{13}$CO core.}
\end{center}

\end{figure}

\begin{table*}
\begin{center}
\caption{Fitted and derived parameters for the ammonia cores.}
\label{tbl:ammonia_fits}
\begin{minipage}{\linewidth}
\begin{tabular}{lcccccccc}
\hline
\hline

Id & Vel. & $T_{\rm{ex}}$ & FWHM & $\tau$ & $N$(H$_2$) & $n$(H$_2$) & Mass & Virial Mass\\
& (km~s$^{-1}$) & (K) & (kms$^{-1}$) &  & (cm$^{-2}$)& (cm$^{-3}$) & (M$_\odot$)& (M$_\odot$)\\

\hline
A & $-$37.0 & 8.0 &  1.63 & 1.41 & 4.0--$7.7\times10^{22}$& 0.7--$1.6\times10^{5}$ & 8--15 & 2.0\\
B & $-$35.3 & 6.4 &  1.23 & 2.58 & 4.3--$8.6\times10^{22}$& 1.5--$3.1\times10^{5}$& 3.5--7 & 0.6\\
\hline
\end{tabular}\\
\\
\end{minipage}
\end{center}
\end{table*}

In Fig.~\ref{fig:ammonia_profiles} we present spectral line profiles integrated over each  ammonia core. The hyperfine components can clearly
be seen in these two spectra and can thus be used to estimate the optical depth for each core using the intensity ratio between the different hyperfine
components. These profiles were fitted using the CLASS software package, in particular the METHOD NH$_3$ utility, assuming that all of the components
have equal excitation temperatures and that the line widths and separations are identical to laboratory values. Once the optical depth is known the
excitation temperature of the ammonia (1,1) inversion can be derived using Eqn.~1 and 2 from \citet{harju1993}

\begin{equation}
T_{\rm{B}}(1,1,m)=\eta\frac{h\nu}{k} \left( F(T_{\rm{ex}})-F(T_{\rm{bg}}) \right) \left(1-\rm{e}^{-\tau(1,1,m)}\right),
\end{equation}

\noindent where the function $F(T)$ is defined as

\begin{equation}
F(T)=\frac{1}{{\rm{e}}^{h\nu/kT}-1}.
\end{equation}

\noindent $T_{\rm{B}}$ is the main beam brightness temperature, $T_{\rm{ex}}$ is the source excitation temperature and $T_{\rm{bg}}$ is the cosmic background temperature assumed to be $\sim2.7$~K, $\eta$ is the beam filling factor, which we assume to be of order unity, and $m$ refers to the main hyperfine group. We have estimated the column density of molecules in the upper (1,1) state using Eqn.~3 from \citet{harju1993},

\begin{equation}
N_{\rm{u}}(1,1)=1.6\times10^{13} F(T_{\rm{ex}})\Delta v~\tau(1,1,m)~\rm{cm}^{-2},
 \end{equation}

\noindent where $\Delta v$ is the FWHM of the main line in km~s$^{-1}$. To obtain the total column density for molecules in the (1,1) state we need to sum the populations of both the upper and lower states. This can be done by applying the Boltzmann equation assuming both the upper and lower levels are evenly populated,

\begin{equation}
N(1,1)=N_{\rm{u}}+N_{\rm{l}}=N_{\rm{u}}\left(1+{\rm{e}}^{h\nu/kT_{\rm{ex}}}\right).
\end{equation}

%A=2.069*2.3
%B=2.0867*2.3

The following equation calculates the total column density $N$(NH$_3$) making the assumption that only metastable levels are populated,

\begin{equation}
N({\rm{NH}}_3)=N(1,1)\left( \frac{1}{3} {\rm{e}}^{23.4/T_{12}}+1+\frac{5}{3}{\rm{e}}^{-41.5/T_{12}}+\frac{14}{3}{\rm{e}}^{-101.5/T_{12}}\right).
\end{equation}

Since there is  no data available for the ammonia (2,2) transition we are unable to calculate the rotational temperature, $T_{12}$. However, we are
able to calculate a range of values for each parameter of interest, such as column and particle densities and core masses, using a range of rotational
temperatures typical of similar cores (10--40~K; e.g., \citealt{wu2006}). This range of temperatures corresponds to a difference of at most a factor of
two in column density and so will not affect our results significantly. In order to obtain H$_2$ column densities we have assumed an ammonia fractional
abundance of $3\times10^{-8}$ (\citealt{wu2006}). The H$_{2}$ densities and core masses have been calculated using the core diameters presented
in Table~\ref{tbl:peak_positions}.  The results of the fits to the ammonia spectral profiles and  core-averaged values for the derived
physical parameters of the cores are presented in Table~\ref{tbl:ammonia_fits}. We consider the masses and densities of the two cores to be lower limits due to the missing short spacings data; taking into account the uncertainties involved in their estimation they are likely to be significantly larger. Both cores appear to be centrally condensed with peak temperatures and densities a factor of $\sim1.5$ and 2 larger, respectively, than the core-averaged values presented in Table~\ref{tbl:ammonia_fits}.  We can test the stability
of these cores by comparing their masses to their virial masses, which we calculate using the standard equation (e.g., \citealt{evans1999}),

\begin{equation}
M_{\rm{vir}}\simeq210R_{\rm{core}} \langle\Delta v\rangle^2,
\end{equation} 

\noindent where $R_{\rm{core}}$ is the ammonia core radius in parsecs and $\Delta v$ is the FWHM line width. The resulting virial masses for each core
are presented in Table~\ref{tbl:ammonia_fits}. Comparing these to the derived core masses reveals that both cores are significantly more massive than could be supported by their internal pressure alone (inferred from their line widths). Therefore these cores are gravitationally bound and given the large difference between their actual masses and their virial masses it is likely they are in a state of gravitational collapse. 

The visual extinction can be estimated using the column density ($N_{\rm{H_2}}$) towards each core using the following relationship: $A_v=N_{\rm{H_2}}/0.94 \times 10^{21}$~mag (\citealt{frerking1982}). This results in similar values for both cores, $A_v\sim 42$--82 and 45--91 for Core~A and Core~B respectively. The visual extinction can be converted to K-band extinction ($A_{K_{\rm{s}}}$) by multiplying by the ratio between visual and infrared extinction i.e.,  $A_{K_{\rm{s}}}=A_v/8.9$ (\citealt{rieke1985}). In this way we obtain K-band extinctions between five and ten magnitudes for both cores. 

\subsection{2MASS and GLIMPSE point sources}
\label{sect:2mass}
\begin{figure*}
\begin{center}
\includegraphics[width=0.49\linewidth]{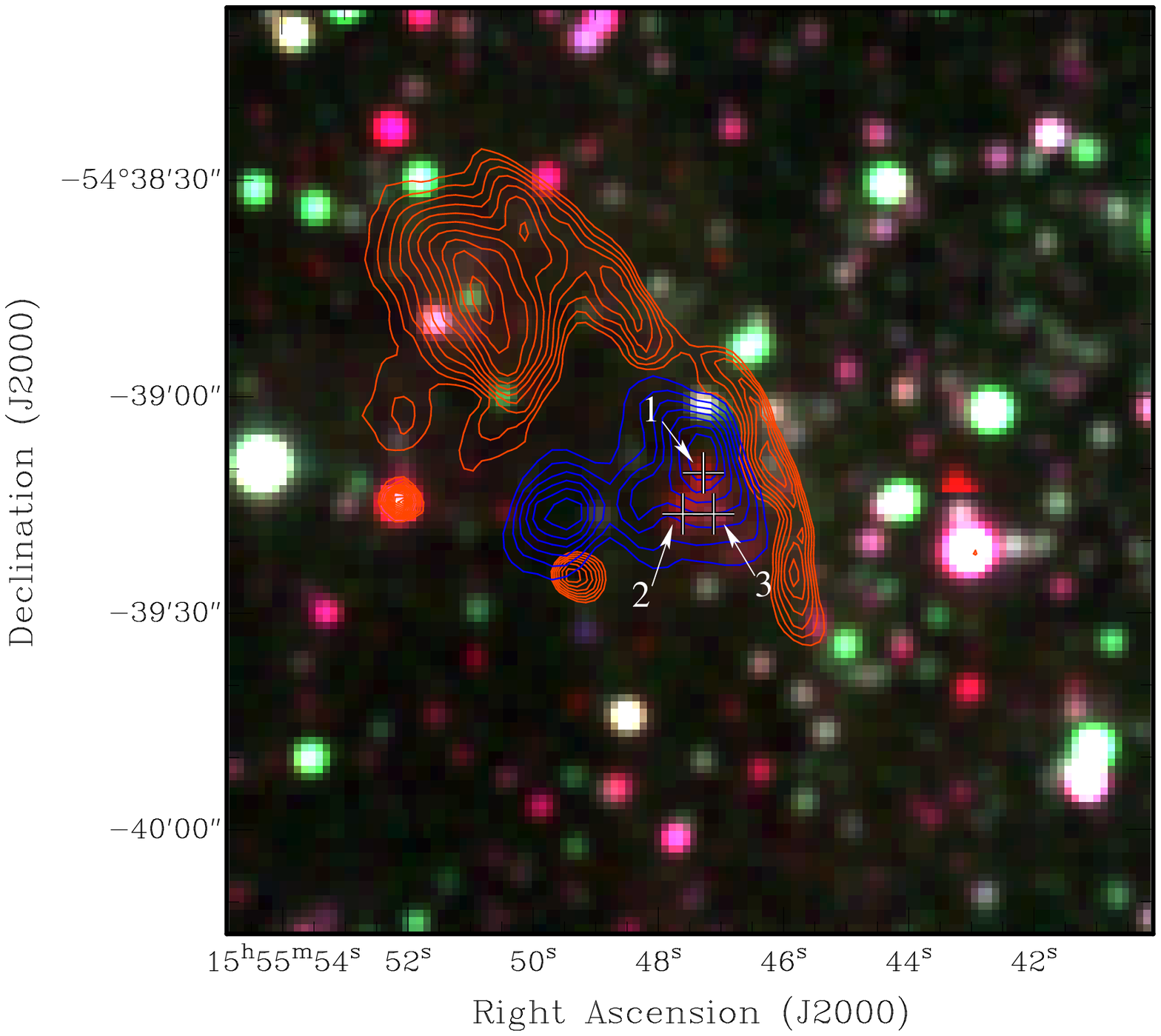} 
\includegraphics[width=0.49\linewidth]{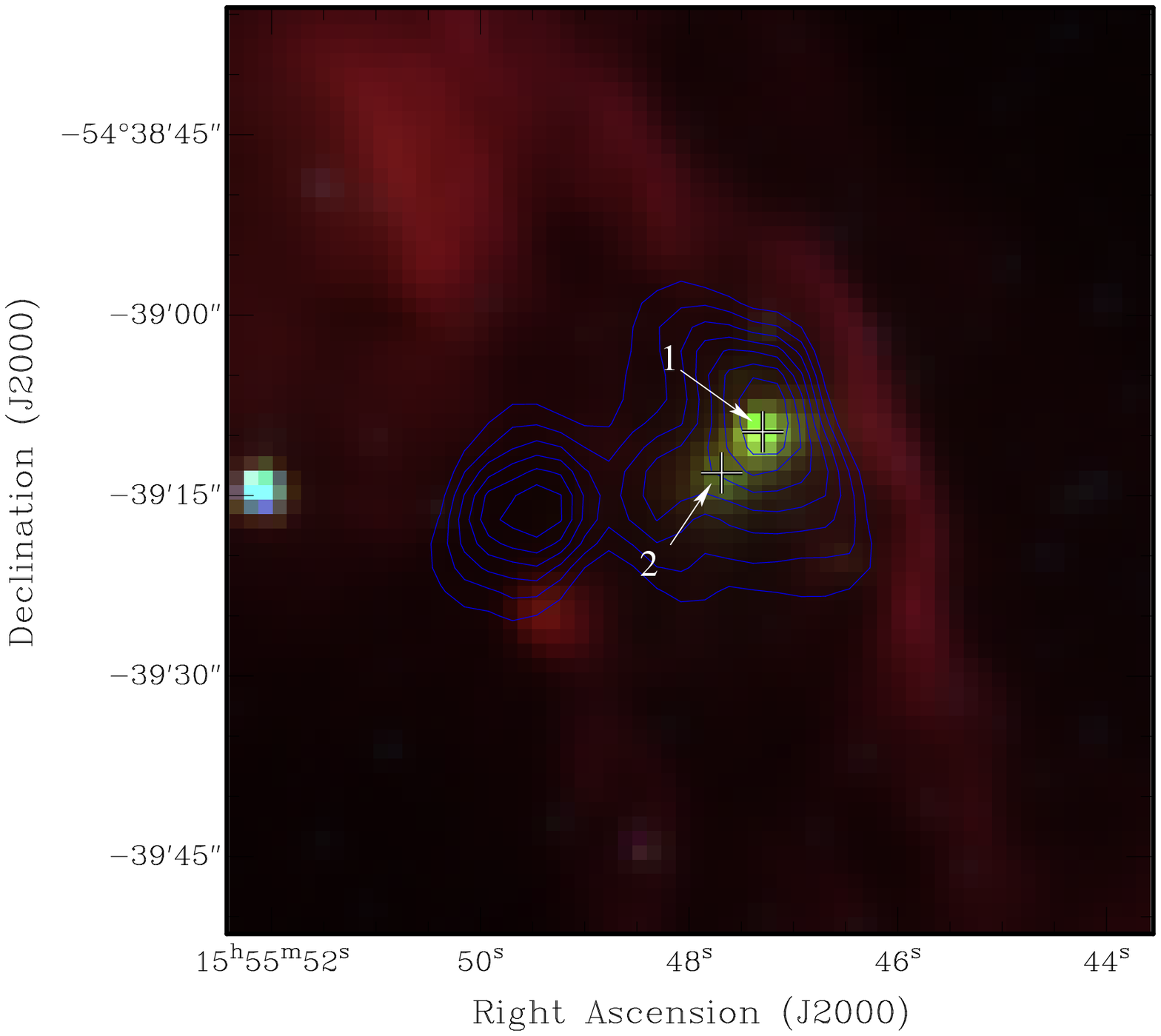}
\caption{\label{fig:2mass} 2MASS and GLIMPSE three colour composite images. Left panel: 2MASS composite produced by combining $J$, $H$ and $K_{\rm{s}}$ band images which are shown in blue, green and red respectively. This image is overlaid with contours of the GLIMPSE 8~$\mu$m emission and the integrated NH$_3$ emission in orange and blue respectively. Right panel: GLIMPSE composite made from the 3.6, 4.5 and 5.8~$\mu$m band images which are coloured blue, green and red respectively. The positions of the 2MASS sources in both images are identified by the reference given in Table~5.}
\end{center}

\end{figure*}

In this subsection we present complementary near- and mid-IR archival data obtained from the 2MASS and GLIMPSE surveys to investigate the stellar content of the ammonia cores. In the \emph{left panel} of Fig.~\ref{fig:2mass} we present a three-colour composite image produced by combining the 2MASS $H$, $J$ and $K_{\rm{s}}$ band images coloured blue, green and red respectively.\footnote{Obtained from \texttt{http://www.ipac.caltech.edu/2mass/.}} To show the spatial relationship between different regions (i.e., the dense cores, PDR and any embedded sources detected in the 2MASS data) this image has been overlaid with contours of the GLIMPSE 8~$\mu$m and the integrated ammonia (1,1) emission coloured orange and blue respectively. This composite image reveals the presence of a small cluster of three very red sources towards the centre of Core~A (these are identified in Fig.~7 by the numbers 1, 2 and 3). None of these 2MASS sources are detected at $J$ band, and only one is detected at $H$ band, which would suggest that they are deeply embedded and that they could be protostellar in nature.
The offsets from the centroid of the core position and their $J$, $H$, $K_{\rm{s}}$ magnitudes and associated quality flags are presented in
Table~\ref{tbl:2mass_data}. 

In the \emph{right panel} of Fig.~\ref{fig:2mass} we present a three-colour composite image of SFO~75 produced by combining the GLIMPSE 3.6~$\mu$m, 4.5~$\mu$m and 5.8~$\mu$m
band images. Contours of the integrated ammonia emission are overlaid in blue. The PDR can clearly be seen as a thin red rim of emission. This image reveals
what appears to be an extended source which is elongated perpendicular to the rim. The peak of this emission correlates with the brightest of the detected 2MASS sources and is extended in the direction of the second brightest 2MASS source (identified as source 1 and 2 respectively in Fig.~7 and Table~5). The green colour is due to a 4.5~$\mu$m excess; this band is
generally dominated by stars, however, diffuse emission has been attributed to a combination of Brackett $\alpha$ and possibly H$_2$ and CO in shocked
regions. Diffuse 4.5~$\mu$m emission has also been associated with molecular outflows -- it is therefore possible that this diffuse emission
is the result of ongoing star formation within Core~A.

\begin{table*}
\begin{minipage}[t]{\linewidth}
\caption{\label{tbl:2mass_data} Parameters of red 2MASS sources located within the contoured boundary of Core~A. The photometric measurements and their associated quality flags have been obtained from the 2MASS Point Source Catalogue (\citealt{cutri2003}).}

\renewcommand{\footnoterule}{}  % to avoid a line before footnotes

\begin{tabular}{cccccccc}
\hline
\hline
Reference 	& Offset& RA & Dec & \multicolumn{3}{c}{2MASS Magnitudes} & Quality\\
		ID			& ($^{\prime\prime}$)& (J2000)&(J2000)&$J$ & $H$ & $K_{\rm{s}}$ & flags\footnote{These flags are as follows: (U) upper limit, (E) very poor fit, (B) SNR $>$ 7 and (A) SNR $>$ 10.} \\
\hline
1 & 2.4 & 15:55:47.3 & -54:39:11.0	& 17.1 & 14.9 & 12.8 &UUB\\
2 & 6.6 & 15:55:47.6 & -54:39:14.8	& 16.8 & 14.4 & 12.8 &UUE\\
3 & 8.0 & 15:55 47.1 & -54:39:16.3	& 16.2 & 14.6 & 13.1 &UBA\\

\hline

\end{tabular}
\end{minipage}
\end{table*}

\begin{figure}
\begin{center}
\includegraphics[width=0.99\linewidth]{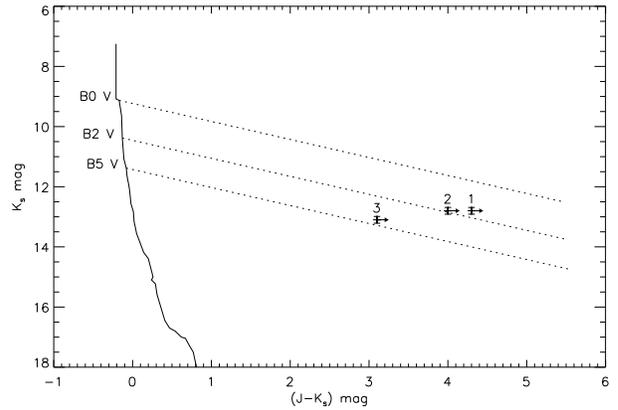} 

\caption{\label{fig:colour_mag} 2MASS \ks\ versus $J-$\ks\ magnitude-colour diagram of the objects detected towards Core~A. The thick black line shows \ks\ magnitudes of main sequence stars located at a distance of 2.8~kpc, taken from \citet{martins2006} and \citet{schmidt-kaler1982}, and assuming zero visual extinction along the line of sight. The dotted lines correspond to a visual extinction of 30~mag, which have been calculated using the standard extinction law of \citet{mathis1990} with $R_{\rm{v}}=3.1$.}
\end{center}

\end{figure}

As previously mentioned, the 2MASS survey was not deep enough to have detected any of these sources at $J$ band and only one source at $H$ band. This limits our analysis of these sources and makes it impossible to use their near-IR photometric colours to determine the evolutionary state of each source (i.e., protostellar, T-Tauri or ZAMS). However, we can use the upper limits quoted for the $J$ band magnitude to derive a lower limit for spectral type and visual extinction towards each source. In Fig.~\ref{fig:colour_mag} we present a \ks\ versus $J-$\ks\ magnitude-colour diagram. Three things are immediately apparent from this plot. First,  two of the objects are of spectral type B2 or earlier with the third source being no later than B5. Secondly, all of the objects are embedded behind $\sim18$--23 magnitudes of visual extinction. This value is approximately half that calculated from the H$_2$ column density through each core (see previous subsection) and places them roughly in the the centre of Core~A. Moreover, the high values of extinction exclude the possibility that these sources are highly reddened main sequence stars located on the far side of the
cloud. Finally, taking account of the 2MASS limiting \ks\ detection limit and assuming an average visual extinction of 20 mag the latest spectral type detectable would be a B5 ($\sim 5.9$~M$_\odot$; \citealt{schmidt-kaler1982}) and therefore the cluster may have many more lower mass members that remain undetected.

The presence of two stars of spectral type B2 or earlier is in line with the findings from our mm-continuum analyses which suggests the presence of either a single high-mass star or a cluster of intermediate-mass stars. Stars with a spectral type of B3 or earlier are known to produce a significant proportion of their radiation as photons with energies high enough to ionise the surrounding material leading to the formation of an ultra-compact HII region. Therefore, we might expect to detect radio emission associated with both of the B2 or earlier stars detected. However, no radio emission has been detected towards any of the embedded sources within Core~A. The non-detection of any radio emission could indicate that the embedded stars are still at a relatively early stage in their evolution, somewhat later than the hot molecular core stage, which are generally not detected at mid-IR wavelengths, but before a detectable HII region has been able to form. This is strongly supported by the detection of a methanol maser coincident with the centre of Core~A recently discovered by the Methanol Multibeam (MMB) survey (MMB Consortium, private comm.)\footnote{For more details see http://www.jb.man.ac.uk/research/methanol/}. Methanol masers are almost exclusively associated with the earliest stages of massive star formation turning off shortly after the formation of the UCHII region (\citealt{walsh1998}). The detection of maser emission towards the centre of Core~A, which has a characteristic lifetime of 2.5--$4.5\times10^4$~yr (\citealt{van_der_walt2005}), allows us to place an upper limit on the age of the star formation taking place within Core~A of no more than $5\times10^4$~years old. We are therefore able to conclude that these sources are deeply embedded massive young stellar objects (MYSOs).

Conversely to Core~A, no 2MASS or GLIMPSE sources are seen towards Core~B. The high values of extinction derived mean that we are limited to detecting only intermediate- to high-mass stars towards both cores. This makes it hard to rule out the presence of ongoing star formation within Core~B completely, however, the non-detection of any stellar source from the 2MASS and GLIMPSE images does allow us to exclude the possibility that any massive star formation is taking place within Core~B. If star formation is taking place within Core~B it is likely to be limited to relatively low-mass stars compared to that taking place in Core~A. Deeper near-IR imaging is necessary before the full stellar inventories of each core can be investigated in detail.

\section{Discussion}

There can be little doubt that the ionisation front is currently having a dramatic influence on the evolution of this cloud. The spatial distributions and morphologies of the IBL, PDR and foremost ammonia core are clearly correlated with one another, strongly suggesting these regions are interacting. In this subsection we will attempt to answer the question of whether the interaction between the ionisation front and the molecular cloud has influenced the observed star formation.

So far we have presented detailed analyses of the structure of the cloud and its associated HII~region. We find compelling evidence that SFO~75 is undergoing RDI by way of an over-pressured IBL and that gas dynamics indicative of shocks are being driven into the surface layers of the cloud. Additionally, we find strong evidence of recent massive star formation within a dense core located at the head of the cloud. The 1.2~mm data reveal the presence of a dense core located behind the cloud's rim. This core appears to be centrally condensed which would imply it is gravitationally bound or contains gravitationally bound objects within it, such as YSOs. The core has a temperature of $\sim30$~K -- significantly higher than found towards starless cores and Bok globules -- which would suggest it is internally heated, further supporting the hypothesis that the core contains protostars that are warming the surrounding material. The high luminosity and mass are similar to those found towards other massive star forming regions (\citealt{sridharan2002}).

The higher angular resolution of the ammonia map resolves the 1.2~mm core into two smaller distinct components which are aligned perpendicular to the bright rim. It is towards the foremost of these two cores that we find strong evidence supporting the conclusions reached from our 1.2~mm data with the detection of a small deeply embedded cluster of high- and intermediate-mass YSOs. Since there is no evidence that star formation is taking place within Core~B, located farther back from the cloud's rim, it would appear that the star formation is taking place exclusively very close to the edge of the rim of the cloud within Core~A. The fact that the star formation is localised to the rim  and is taking place within a core that is being strongly affected by the ionisation front and its associated shocks is persuasive circumstantial evidence that the star formation has been triggered. 

The non-detection of any star formation activity towards Core~B may indicate that while the ionisation front has had a dramatic affect on Core~A, its associated shock front has not yet propagated far enough into the cloud to have had a significant effect on the kinematics of Core~B. In this regard it is interesting to note that Core~A is marginally warmer, more massive, and has wider line widths than Core~B, in line with what would be expected if it is undergoing RDI. Moreover, we note that Core~B appears to have a much simpler, almost spherical, morphology and is offset in velocity by approximately 2 km~s$^{-1}$ from that of Core~A. This situation is similar to examples of \emph{small scale sequential star formation} seen towards other BRCs (e.g.,~\citealt{sugitani1995,ogura2002,ogura2006}).

This hypothesis is supported by our analysis of the shocked layer identified from the CO data presented in Sect.~\ref{sect:co_radio} which estimated the shock had only propagated as far as the mid-point between the two ammonia cores. If the blue and red shifted emission seen in the $^{13}$CO position-velocity
diagram is due to the propagation of an ionisation induced shock front, we
would expect the time taken for the shock to have propagated to this point to be
similar to the time that the cloud has been exposed to UV illumination from
LSS 3423. The shock propagation speed can be estimated using the relationship between the pre- and post-shock pressures of the neutral gas (i.e., \citealt{white1999}),

\begin{equation}
V_{\rm{s}}^2=\alpha \frac{(P_{\rm{s}}-P_{\rm{n}})}{\rho_{\rm{n}}},
\end{equation}

\noindent where $P_{\rm{s}}$ is the pressure of the shocked gas, $P_{\rm{n}}$ and  $\rho_{\rm{n}}$ are the pressure and density in the pre-shocked neutral gas respectively. The pressure of the shocked gas has been assumed to be twice that of the ionised surface layer. The value of $\alpha$ lies between 1 and 2 and using White et al's assumption that the pre- and post-shocked density ratio has a range between \mbox{2--$\infty$}, results in a maximum error of $\sqrt 2$ into the estimated shock velocity. Using these assumptions we derive a shock velocity of $\sim2$~\kms; this corresponds to a propagation time for the shock of $\sim1.6\times 10^5$~yr.

We can estimate the amount of time the cloud has been exposed to the ionisation front by first calculating the age of the HII region and then comparing it to the time taken for the HII region to have expanded to the edge of the cloud. The boundary between the ionised gas and the surrounding molecular gas is clearly defined by the PDR which separates the two regions. We have used this interface to estimate the angular radius of the HII region to be $\sim 3.5$\arcmin, corresponding to a physical radius of 2.8~pc. When the ionising star turns on, its radiation begins ionising the surrounding medium leading to the creation of a HII region that rapidly expands out to the Str\"omgren radius, after which point it continues to expand more slowly. Taking account of the Str\"omgren radius for a O9.5 star (0.3~pc and 1.5~pc for a HII~region expanding into a medium of densities of 10$^3$~\cmthree\ and 10$^2$~\cmthree\ respectively),
and using an expansion speed of 11.4~\kms\ for the period of expansion after the Str\"omgren radius has been reached, we estimate the age of the HII
region to be approximately 1--$2.5\times 10^5$~yr. At a projected distance of 0.8~pc the amount of time SFO~75 has been exposed to the ionisation radiation ranges between 1--$2\times 10^5$~yr.

Comparing the ionisation time to the shock propagation time we find them to be
of similar age. Although this is not conclusive proof that the shock is directly connected to the arrival of the ionising front at the surface of the
cloud, our simplistic analysis does not rule out a connection and shows them to be approximately coeval. Given that the star formation almost certainly post-dates the arrival of the ionisation front, the strong evidence of shocks within the surface layers of the cloud, and the excellent morphological correlation between the IBL, PDR and flattered face of Core~A we conclude that the star formation has been triggered by the photoionisation-induced shock driven into the cloud by the expanding HII region. The lack of any star formation activity within Core~B further strengthens the case for the star formation within Core~A to have been triggered.

\section{Summary and conclusions}
\label{sect:summary}

We present a multi-wavelength study of the bright-rimmed cloud SFO~75 in order to evaluate whether or not the observed star formation could have been
triggered by radiative-driven implosion. This study includes observations of the 1.3~cm and 1.2~mm continuum, and $^{13}$CO ($J$=1--0) and ammonia (1,1) spectral line observations. To complement our observations we have used near-, mid- and far-IR archival data obtained from the
2MASS, GLIMPSE, and IRAS archives respectively. Combining all of these data sets has enabled us to build up a comprehensive picture of this cloud and
its local environment. 

RCW~98 appears to be an approximately spherical HII~region surrounded by a low density ($< 10^3$~\cmthree) ring of molecular material. The HII region has an angular radius of $\sim3\arcmin.5$, corresponding to a physical radius of $\sim2.8$~pc, from which we estimate its age to be $1$--$2.5\times10^5$~yr.
SFO~75 is located on the south-east side of the HII~region, and approximately lies in the same plane in the sky as the ionising star. The bright rim of SFO~75 is located $\sim0.8$~pc from the ionising star and has been exposed to the ionising radiation for $\sim1$--$2\times10^5$~yr. 

We are able to clearly identify the ionised boundary layer (IBL), the photo-dominated region and the molecular material. The excellent morphological correlation between these three leads us to conclude that there is a high degree of interaction between them. The IBL is over-pressured with respect to the molecular gas by a factor of three and it is therefore likely that photoionisation-induced shocks are currently being driven into this cloud. This is strongly supported by the detection of red and blue shifted emission which is attributed to emission from shocked gas. We estimated the shock propagation time to be $\sim 1.6\times10^5$~yr and find it to be approximately coeval with the cloud's exposure to the HII~region.

Our CO and 1.2~mm data reveal the presence of a dense molecular core located slightly behind the bright rim. The density and mass of the molecular gas range between $\sim0.3$--$3\times10^5$~\cmthree\ and 170--570~M$_\odot$ respectively.
We find strong evidence of recent massive star formation within a dense core located at the head of the cloud. This core appears to be centrally condensed which would imply it is gravitationally bound or contains a gravitationally bound object within it. The core temperature ($\sim30$~K) is significantly higher than found towards starless cores and Bok globules indicating the presence of an  internal source of  heating, such as YSOs. Moreover, the high luminosity and mass are similar to those found towards other massive star forming regions (\citealt{sridharan2002}).

The 1.2~mm core is resolved by our high resolution ammonia observations into
two distinct cores, Core~A which is located directly behind the bright rim and Core~B which is located farther back. Towards Core~A we find a small cluster of MYSOs (two B2 and one B5 stars or earlier) which are deeply embedded behind $\sim20$~mag of visual extinction. The upper limit for the age of this cluster, which is approximately half the estimated ionisation time, would indicate that its formation post-dates the cloud's exposure to the HII~region. Due to the limited magnitude of the 2MASS observations we are unable to detect any sources that might be present that are later than B5, so these sources might just be the most massive members of a larger cluster that remains hidden. No stars are found towards Core~B and given its simple morphology and that it is located farther back from the rim than Core~A leads us to conclude that the shock front has not yet penetrated far enough into the cloud to have reached this core.

Although we are unable to determine conclusively whether the observed star formation has been triggered we have presented a considerable amount of circumstantial evidence which we believe strongly supports the hypothesis that the star formation has been induced. The scenario that emerges from our analysis is one where the two ammonia cores pre-date the arrival of the ionisation front. However, since its arrival the over-pressure of the ionised gas at the surface of the cloud has driven shocks into the surface layers of the cloud. The propagation of these shocks through Core~A has triggered the formation of a small cluster of massive stars, however, the shock front has not yet propagated deeply enough into the cloud to have affected the evolution of Core~B.

\begin{acknowledgements}

The authors would like to thank the Director and staff of the Paul Wild Observatory, Narrabri, New South Wales, Australia for their hospitality and
assistance during our observations at the Compact Array and Mopra Telescope. JSU is supported by a PPARC postdoctoral grant. This research
would not have been possible without the SIMBAD astronomical database service operated at CDS, Strasbourg, France and the NASA Astrophysics Data System
Bibliographic Services. This research makes use of data products from the Two Micron All Sky Survey, which is a joint project of the University of
Massachusetts and the Infrared Processing and Analysis Center/California Institute of Technology, funded by the National Aeronautics and Space
Administration and the National Science Foundation.

\end{acknowledgements}

\bibliography{7236bib}

\bibliographystyle{aa}
\end{document}